\documentclass[11pt,ams]{article}
\usepackage{graphicx}
\usepackage{amsmath}

\begin{document}

\date{}
\title{\textbf{Local renormalizable gauge theories from nonlocal operators }}
\author{\textbf{M.A.L. Capri}$^a$\thanks{
marcio@dft.if.uerj.br} \ , \textbf{V.E.R. Lemes}$^a$\thanks{
vitor@dft.if.uerj.br} \ , \textbf{R.F. Sobreiro}$^b$\thanks{
sobreiro@cbpf.br} \and \textbf{S.P. Sorella}$^a$\thanks{
sorella@uerj.br}{\ }\ , \textbf{R. Thibes}$^a$\thanks{
thibes@dft.if.uerj.br} \\
[3mm]\textit{$^a$ UERJ $-$ Universidade do Estado do Rio de Janeiro} \\
\textit{Rua S{\~a}o Francisco Xavier 524, 20550-013, Maracan{\~a}}, \\
\textit{Rio de Janeiro, Brasil} \\
[3mm]\textit{$^b$ CBPF, Centro Brasileiro de Pesquisas F{\'\i}sicas} \\
\textit{Rua Xavier Sigaud 150, 22290-180, Urca}, \\
\textit{Rio de Janeiro, Brasil}}
\maketitle

\begin{abstract}
The possibility that nonlocal operators might be added to the Yang-Mills
action is investigated. We point out that there exists a class of nonlocal
operators which lead to renormalizable gauge theories. These operators turn
out to be localizable by means of the introduction of auxiliary fields. The
renormalizability is thus ensured by the symmetry content exhibited by the
resulting local theory. The example of the nonlocal operator $Tr$\textbf{$%
\int A_{\mu }\frac{1}{D^{2}}A_{\mu }$ }is analysed in detail. A few remarks
on the possible role that these operators might have for confining theories
are outlined.
\end{abstract}

\newpage


\section{Introduction}

The understanding of the behavior of Yang-Mills theories in the
nonperturbative infrared region is a great challenge in quantum field
theory. Different approaches are currently employed to address this issue,
namely: lattice gauge theories\footnote{%
See ref.\cite{Greensite:2003bk} for a general overview.}, study of the
Schwinger-Dyson equations \cite
{Alkofer:2004it,vonSmekal:1997is,vonSmekal:1997vx,Alkofer:2000wg}, duality
mechanisms \cite
{Maldacena:1998im,Polchinski:2001tt,Boschi-Filho:2005yh,Brodsky:2006uq,Andreev:2007vn}%
, restriction of the domain of integration in the Feynman path integral in
order to take into account the existence of the Gribov copies \cite
{Gribov:1977wm,Zwanziger:1989mf,Zwanziger:1992qr}, variational principles
\cite{Reinhardt:2004mm,Feuchter:2004mk}, condensates \cite
{Shifman:1978bx,Shifman:1978by,Gubarev:2000eu,Gubarev:2000nz,Verschelde:2001ia,Dudal:2003vv,Dudal:2002pq,Gracey:2004bk,Dudal:2003by,Dudal:2004rx}%
, exact renormalization group \cite{Pawlowski:2003hq,Fischer:2006vf}.
Several results have been achieved so far, having received confirmation from
the various approaches. This is the case, for example, of the infrared
suppression of the two point gluon correlation function and of the infrared
enhancement of the ghost propagator in the Landau gauge \cite
{Gribov:1977wm,Zwanziger:1989mf,Zwanziger:1992qr,Alkofer:2004it,vonSmekal:1997is, vonSmekal:1997vx,Alkofer:2000wg,Pawlowski:2003hq,Fischer:2006vf, Cucchieri:2004mf,Bloch:2003sk,Cucchieri:1997dx,Furui:2006ks,Furui:2006rx, Bogolubsky:2005wf,Bowman:2007du}%
. Nevertheless, a satisfactory description of the gluon and quark
confinement is not yet at our disposal. One still has the feeling that much
work is needed. \newline
\newline
The aim of this paper is to call attention to the fact that there exist
nonlocal operators which can be consistently added to the Yang-Mills action.
This means that, for those specific operators, a renormalizable
computational framework can be worked out. As is well known, adding a
nonlocal term to the Yang-Mills action is a delicate operation. In most
cases the requirement of renormalizability cannot be accomplished. However,
in a few cases, the nonlocal term can be cast in local form through the
introduction of additional localizing fields. Furthermore, the resulting
local theory might exhibit a rich content of symmetries, enabling us to
establish its multiplicative renormalizability to all orders. It is worth
underlining that, being nonlocal, these operators can induce deep
modifications on the large distance behavior of the theory. As such, they
might be useful in order to investigate nonperturbative features, being of
particular interest for confining theories. \newline
\newline
As an explicit example of such nonlocal terms, we shall present a detailed
analysis of the nonlocal operator
\begin{equation}
\mathcal{O}=\frac{1}{2}\int d^{4}\!x\,A_{\mu }^{a}\left( \frac{1}{D^{2}}%
\right) ^{ab}A_{\mu }^{b}\;,  \label{i0}
\end{equation}
where $D^{2}$ stands for the covariant Laplacian
\begin{eqnarray}
\left( D^{2}\right) ^{ab} &=&D_{\mu }^{ac}D_{\mu }^{cb}\;,  \nonumber \\
D_{\mu }^{ab} &=&\delta ^{ab}\partial _{\mu }-gf^{abc}A_{\mu }^{c}\;.
\label{i01}
\end{eqnarray}
Through this example we shall be able to provide a general overview of what
can be called a consistent framework \cite{Capri:2005dy,Capri:2006ne} for a
nonlocal operator which can be added to the Yang-Mills action, namely:

\begin{itemize}
\item  achievement of a localization procedure,

\item  investigation of the symmetry content of the resulting local action,

\item  proof of the multiplicative renormalizability of the theory.
\end{itemize}

\noindent In order to have an idea of the relevance of such nonlocal terms
for the infrared behavior of Yang-Mills theories, let us spend a few words
on two examples which have been analysed recently, and which fulfil the
requirements of localizability and renormalizability. The first example is
provided by the Zwanziger horizon term which implements the restriction of
the domain of integration in the Feynman path integral to the Gribov region $%
\Omega $ in the Landau gauge\footnote{%
For the generalization of the horizon function $\left( \ref{i1}\right) $ to
the maximal Abelian gauge see \cite{Capri:2005tj,Capri:2006cz}.} \cite
{Zwanziger:1989mf,Zwanziger:1992qr}, namely
\begin{equation}
S_{\mathrm{H}}=-g^{2}\gamma ^{4}\int d^{4}x\;f^{abc}A_{\mu }^{b}\left( \frac{%
1}{\left( \partial _{\nu }D_{\nu }\right) }\right) ^{ad}f^{dec}A_{\mu
}^{e}\;,  \label{i1}
\end{equation}
where the parameter $\gamma $, known as the Gribov parameter, has the
dimension of a mass. The second example is given by the gauge invariant
nonlocal operator
\begin{equation}
S_{\mathrm{m}}=\frac{m^{2}}{2}\int d^{4}xF_{\mu \nu }^{a}\left( \frac{1}{%
D^{2}}\right) ^{ab}F_{\mu \nu }^{b}\;,  \label{i2}
\end{equation}
which, when added to the Yang-Mills action, yields an effective
gauge invariant mass $m$ for the gluons
\cite{Jackiw:1997jg,Capri:2005dy,Capri:2006ne}, a topic which is
receiving increasing attention in recent years. As shown in \cite
{Zwanziger:1989mf,Zwanziger:1992qr,Maggiore:1993wq,Dudal:2005na,Capri:2005dy,Capri:2006ne}%
, both operators $\left( \ref{i1}\right) $,$\left( \ref{i2}\right) $ are
localizable, the resulting local theories enjoy the property of being
renormalizable. In particular, in \cite
{Gracey:2006dr,Gracey:2007vv,Capri:2006ne} one finds the two loop
calculation of the anomalous dimensions corresponding to expressions $\left(
\ref{i1}\right) $,$\left( \ref{i2}\right) $. \newline
\newline
Concerning now the operator $\mathcal{O}$, eq.$\left( \ref{i0}\right) $, a
few potential interesting features might be pointed out in order to motivate
better its investigation. We observe that its introduction in the Yang-Mills
action leads to a deep modification of the gluon propagator in the infrared.
More precisely, as we shall see in the next section, the addition of the
term $\left( \ref{i0}\right) $ will give rise to a tree level gluon
propagator which is of the Gribov type \cite
{Gribov:1977wm,Zwanziger:1989mf,Zwanziger:1992qr}, \textit{i.e. }it is
suppressed in the infrared, exhibiting positivity violation, a feature
usually interpreted as a signal of confinement. This should be not
surprising. Notice in fact that, in the quadratic approximation, both
operators $\left( \ref{i0}\right) $,$\left( \ref{i1}\right) $ reduce to the
same expression, thus yielding the same propagator. Also, we mention that
expression $\left( \ref{i0}\right) $ can be easily adapted to the lattice
formulation, thus it could also be investigated through numerical
simulations. \newline
\newline
The present work is organized as follows. In section 2 we describe the
localization procedure for the operator $\left( \ref{i0}\right) $. Section 3
is devoted to the study of the symmetry content of the resulting local
action. In section 4 we derive the set of Ward identities. In section 5 we
present the algebraic characterization of the most general local invariant
counterterm, and we establish the renormalizability of the model. Section 6
collects the conclusion.

\section{The localization procedure}

Let us start by considering the gauge fixed Yang-Mills action with the
addition of the nonlocal operator $\mathcal{O}$, eq.$\left( \ref{i0}\right) $%
, namely
\begin{equation}
S=S_{\mathrm{YM}}+S_{\mathrm{gf}}+\sigma ^{4}\mathcal{O}\;,  \label{s21}
\end{equation}
where $S_{\mathrm{YM}}$ is the Yang-Mills action in four dimensional
Euclidean space-time,
\begin{equation}
S_{\mathrm{YM}}=\frac{1}{4}\int d^{4}\!x\,F_{\mu \nu }^{a}F_{\mu \nu }^{a}\;,
\label{s22}
\end{equation}
with
\begin{equation}
F_{\mu \nu }^{a}=\partial _{\mu }A_{\nu }^{a}-\partial _{\nu }A_{\mu
}^{a}+gf^{abc}A_{\mu }^{b}A_{\nu }^{c}\;.  \label{s23}
\end{equation}
The term $S_{\mathrm{gf}}$ stands for the gauge fixing term, here taken in
the Landau gauge, \textit{i.e. }
\begin{equation}
S_{\mathrm{gf}}=\int d^{4}\!x\,\left( b^{a}\partial _{\mu }A_{\mu }^{a}+\bar{%
c}^{a}\partial _{\mu }D_{\mu }^{ab}c^{b}\right) \;,  \label{s24}
\end{equation}
where the auxiliary field $b^{a}$ is the Lagrange multiplier enforcing the
Landau gauge condition, $\partial _{\mu }A_{\mu }^{a}=0$, and $(\bar{c}%
^{a},c^{a})$ are the Faddeev-Popov ghost fields. Notice that, in order to
have the correct dimensions, a parameter $\sigma $ with the dimension of a
mass has been introduced in expression $\left( \ref{s21}\right) $. As the
purpose of the present work is that of showing that a local and
renormalizable action can be obtained from the nonlocal expression $\left(
\ref{i0}\right) $, $\sigma $ will be treated as a free parameter. After
having proven the renormalizability of the resulting local theory, one can
address the issue of whether $\sigma $ could be generated in a dynamical
way, being associated to a possible condensation of the operator $\mathcal{O}
$, \textit{i.e.} $\left\langle \mathcal{O}\right\rangle \neq 0$. As it
happens in the case of the Gribov parameter $\gamma $ \cite
{Gribov:1977wm,Zwanziger:1989mf,Zwanziger:1992qr} of the horizon function,
eq.$\left( \ref{i1}\right) $, this would demand that the parameter $\sigma $
is a solution of a suitable gap equation, enabling us to express it as a
function of the gauge coupling constant $g$ and of the invariant scale $%
\Lambda _{QCD}$. Although being out of the aim of the present work, we shall
come back to this interesting point in the conclusion, where a possible
strategy to face the dynamical generation of the parameter $\sigma $ will be
outlined. \newline
\newline
An interesting feature of the action $\left( \ref{s21}\right) $ is that it
gives rise to a tree level gluon propagator which displays the
characteristic Gribov behavior \cite
{Gribov:1977wm,Zwanziger:1989mf,Zwanziger:1992qr}, namely
\begin{equation}
\langle A_{\mu }^{a}(k)A_{\nu }^{b}(-k)\rangle =\delta ^{ab}\frac{k^{2}}{%
k^{4}+\sigma ^{4}}\left( \delta _{\mu \nu }-\frac{k_{\mu }k_{\nu }}{k^{2}}%
\right) \;.  \label{s25}
\end{equation}
As already remarked, this is a consequence of the fact that both operators $%
\left( \ref{i0}\right) ,\left( \ref{i1}\right) $ reduce to the same
expression in the quadratic approximation, thus leading to same the tree
level propagator. \newline
\newline
Although the operator $\left( \ref{i0}\right) $ is nonlocal, it can be cast
in local form by introducing a suitable set of auxiliary localizing fields.
This is performed according to
\begin{eqnarray}
e^{-\sigma ^{4}\mathcal{O}} &=&\int D\bar{B}DBD\bar{G}DG\,\exp \left\{ -\int
d^{4}\!x\,\left[ \frac{1}{2}\left( \bar{B}_{\mu }^{a}D_{\nu }^{ab}D_{\nu
}^{bc}B_{\mu }^{c}-\bar{G}_{\mu }^{a}D_{\nu }^{ab}D_{\nu }^{bc}G_{\mu
}^{c}\right) +\frac{\sigma ^{2}}{2}\left( B-\bar{B}\right) _{\mu }^{a}A_{\mu
}^{a}\right] \right\} \;,  \nonumber \\
&&  \label{loc1}
\end{eqnarray}
where $(\bar{B}_{\mu }^{a},B_{\mu }^{a})$ are bosonic vector fields, while $(%
\bar{G}_{\mu }^{a},G_{\mu }^{a})$ are anticommuting fields. Thus, for the
partition function $\mathcal{Z}$ of the model we may write
\begin{equation}
\mathcal{Z=}\int DADbDcD\overline{c}\;e^{-S}=\int DADbDcD\overline{c}D\bar{B}%
DBD\bar{G}DG\;e^{-S_{\mathrm{Local}}}\;,  \label{pf}
\end{equation}
where the local action $S_{\mathrm{Local}}$ is now given by
\begin{equation}
S_{\mathrm{Local}}=S_{\mathrm{YM}}+S_{\mathrm{gf}}+S_{\mathrm{aux}%
}+S_{\sigma }\;,  \label{local}
\end{equation}
with
\begin{eqnarray}
S_{\mathrm{aux}} &=&\frac{1}{2}\int d^{4}\!x\,\left( \bar{B}_{\mu
}^{a}D_{\nu }^{ab}D_{\nu }^{bc}B_{\mu }^{c}-\bar{G}_{\mu }^{a}D_{\nu
}^{ab}D_{\nu }^{bc}G_{\mu }^{c}\right) \;,  \nonumber \\
S_{\sigma } &=&\frac{\sigma ^{2}}{2}\int d^{4}\!x\,\left( B-\bar{B}\right)
_{\mu }^{a}A_{\mu }^{a}\;.  \label{local2}
\end{eqnarray}

\section{Symmetry content}

To analyze the symmetry content of our model we shall start first by
considering the case in which the parameter $\sigma $ is set to zero,
\textit{i.e.} $\sigma =0$, yielding
\begin{equation}
S_{0}\equiv S_{\mathrm{Local}}|_{\sigma =0}=S_{\mathrm{YM}}+S_{\mathrm{gf}%
}+S_{\mathrm{aux}}\;.  \label{massless}
\end{equation}
The action $\left( \ref{massless}\right) $ is completely equivalent to the
Yang-Mills action, since the introduction of the auxiliary fields $(\bar{B}%
_{\mu }^{a},B_{\mu }^{a})$ and $(\bar{G}_{\mu }^{a},G_{\mu }^{a})$ amounts
simply to inserting a unity factor, \textit{i.e.}
\begin{equation}
1=\int D\bar{B}DBD\bar{G}DG\,\,\exp \left[ -\frac{1}{2}\int d^{4}\!x\,\left(
\bar{B}_{\mu }^{a}D_{\nu }^{ab}D_{\nu }^{bc}B_{\mu }^{c}-\bar{G}_{\mu
}^{a}D_{\nu }^{ab}D_{\nu }^{bc}G_{\mu }^{c}\right) \right] \;.  \label{un}
\end{equation}
Furthermore, the action $\left( \ref{massless}\right) $ enjoys the following
symmetries

\begin{itemize}
\item  {The $BRST$ symmetry:
\begin{eqnarray}
sA_{\mu }^{a} &=&-D_{\mu }^{ab}c^{b}\;,  \nonumber \\
sc^{a} &=&\frac{g}{2}f^{abc}c^{a}c^{b}\;,  \nonumber \\
s\bar{c}^{a} &=&b^{a}\;,  \nonumber \\
sb^{a} &=&0\;,  \nonumber \\
sB_{\mu }^{a} &=&gf^{abc}c^{b}B_{\mu }^{c}\;,  \nonumber \\
s\bar{B}_{\mu }^{a} &=&gf^{abc}c^{b}\bar{B}_{\mu }^{c}\;,  \nonumber \\
sG_{\mu }^{a} &=&gf^{abc}c^{b}G_{\mu }^{c}\;,  \nonumber \\
s\bar{G}_{\mu }^{a} &=&gf^{abc}c^{b}\bar{G}_{\mu }^{c}\;.  \label{brst}
\end{eqnarray}
}

\item  {The $\delta $-symmetry:
\begin{eqnarray}
\delta B_{\mu }^{a} &=&G_{\mu }^{a}\;,  \nonumber \\
\delta G_{\mu }^{a} &=&0\;,  \nonumber \\
\delta \bar{G}_{\mu }^{a} &=&\bar{B}_{\mu }^{a}\;,  \nonumber \\
\delta \bar{B}_{\mu }^{a} &=&0\;.  \label{delta}
\end{eqnarray}
Evidently, }\textit{\ }
\end{itemize}

\begin{equation}
sS_{0}=\delta S_{0}=0\;.  \label{invs0}
\end{equation}
Both operators $s$ and $\delta $ are nilpotent, obeying the following
anticommutation relations
\begin{equation}
s^{2}=\delta ^{2}=\{s,\delta \}=0\;.  \label{ant}
\end{equation}
In the same way as the $BRST$ transformations allow us to introduce the
ghost number operator $\mathcal{N}_{\mathrm{gh}}$%
\begin{equation}
\mathcal{N}_{\mathrm{gh}}S_{0}=\int d^{4}\!x\,\left( c^{a}\frac{\delta }{%
\delta c^{a}}-\bar{c}^{a}\frac{\delta }{\delta \bar{c}^{a}}\right) S_{0}=0\;,
\label{gh}
\end{equation}
the $\delta $-transformations $\left( \ref{delta}\right) $ enable us to
introduce a second operator $\mathcal{N}_{\mathrm{f}\text{ }}$ associated to
the anticommuting fields $(\bar{G}_{\mu }^{a},G_{\mu }^{a})$, namely
\begin{equation}
\mathcal{N}_{\mathrm{f}}S_{0}=\int d^{4}\!x\,\left( G_{\mu }^{a}\frac{\delta
}{\delta G_{\mu }^{a}}-\bar{G}_{\mu }^{a}\frac{\delta }{\delta \bar{G}_{\mu
}^{a}}\right) S_{0}=0\;.  \label{nf}
\end{equation}
Let us now try to take into account the term $S_{\sigma }$ in $\left( \ref
{local2}\right) $. As is easily seen, this term breaks both $s$ and $\delta $
symmetries, in fact
\begin{eqnarray}
sS_{\sigma } &=&-\frac{\sigma ^{2}}{2}\int d^{4}x\left( B-\bar{B}\right)
_{\mu }^{a}\partial _{\mu }c^{a}\;,  \nonumber \\
\delta S_{\sigma } &=&\frac{\sigma ^{2}}{2}\int d^{4}x\;G_{\mu }^{a}A_{\mu
}^{a}\;.  \label{br}
\end{eqnarray}
Moreover, the breaking terms $\left( \ref{br}\right) $ can be kept under
control by embedding the action $\left( \ref{local}\right) $ in a more
general model with exact invariance, a strategy already successfully
employed in the case of Zwanziger's horizon function $S_{\mathrm{H}}$ \cite
{Zwanziger:1989mf,Zwanziger:1992qr}, eq.$\left( \ref{i1}\right) $, and of
the nonlocal mass term $S_{\mathrm{m}}$ \cite{Capri:2005dy,Capri:2006ne}, eq.%
$\left( \ref{i2}\right) $. To this purpose we introduce a set of external
sources, namely
\begin{equation}
\left\{ X_{\mu \nu },\bar{X}_{\mu \nu };Y_{\mu \nu },\bar{Y}_{\mu \nu
};U_{\mu \nu },\bar{U}_{\mu \nu };V_{\mu \nu },\bar{V}_{\mu \nu }\right\} \;,
\label{sex}
\end{equation}
which enable us to introduce the composite operators $B_{\mu }^{a}A_{\nu
}^{a}$, $\bar{B}_{\mu }^{a}A_{\nu }^{a}$, $G_{\mu }^{a}A_{\nu }^{a}$, and $%
\bar{G}_{\mu }^{a}A_{\nu }^{a}$. Requiring that the sources $X_{\mu \nu },%
\bar{X}_{\mu \nu },Y_{\mu \nu },\bar{Y}_{\mu \nu },U_{\mu \nu },\bar{U}_{\mu
\nu },V_{\mu \nu },\bar{V}_{\mu \nu }$ transform as
\begin{eqnarray}
sX_{\mu \nu } &\!\!\!=\!\!\!&Y_{\mu \nu }\;,\qquad sY_{\mu \nu }=0\;,
\nonumber \\
s\bar{Y}_{\mu \nu } &\!\!\!=\!\!\!&\bar{X}_{\mu \nu }\;,\qquad s\bar{X}_{\mu
\nu }=0\;,  \nonumber \\
sV_{\mu \nu } &\!\!\!=\!\!\!&U_{\mu \nu }\;,\qquad sU_{\mu \nu }=0\;,
\nonumber \\
s\bar{U}_{\mu \nu } &\!\!\!=\!\!\!&\bar{V}_{\mu \nu }\;,\qquad s\bar{V}_{\mu
\nu }=0\;,  \label{brst-sources}
\end{eqnarray}
and
\begin{eqnarray}
\delta V_{\mu \nu } &\!\!\!=\!\!\!&-X_{\mu \nu }\;,\qquad \delta X_{\mu \nu
}=0\;,  \nonumber \\
\delta \bar{V}_{\mu \nu } &\!\!\!=\!\!\!&\bar{X}_{\mu \nu }\;,\qquad \delta
\bar{X}_{\mu \nu }=0\;,  \nonumber \\
\delta U_{\mu \nu } &\!\!\!=\!\!\!&Y_{\mu \nu }\;,\qquad \delta Y_{\mu \nu
}=0\;,  \nonumber \\
\delta \bar{U}_{\mu \nu } &\!\!\!=\!\!\!&-\bar{Y}_{\mu \nu }\;,\qquad \delta
\bar{Y}_{\mu \nu }=0\;,  \label{delta-sources}
\end{eqnarray}
it is apparent that the action $\widetilde{S}_{\sigma }$
\begin{eqnarray}
\widetilde{S}_{\sigma } &=&s\delta \int d^{4}\!x\,\left( \bar{U}_{\mu \nu
}B_{\mu }^{a}A_{\nu }^{a}-V_{\mu \nu }\bar{G}_{\mu }^{a}A_{\nu }^{a}\right)
\nonumber \\
&=&\int d^{4}\!x\,\left( -\bar{X}_{\mu \nu }B_{\mu }^{a}A_{\nu }^{a}+Y_{\mu
\nu }\bar{G}_{\mu }^{a}A_{\nu }^{a}+U_{\mu \nu }\bar{B}_{\mu }^{a}A_{\nu
}^{a}+\bar{V}_{\mu \nu }G_{\mu }^{a}A_{\nu }^{a}\right.  \nonumber \\
&&\left. -\left( \bar{Y}_{\mu \nu }B_{\mu }^{a}-X_{\mu \nu }\bar{G}_{\mu
}^{a}-V_{\mu \nu }\bar{B}_{\mu }^{a}-\bar{U}_{\mu \nu }G_{\mu }^{a}\right)
\partial _{\nu }c^{a}\right) \;,  \label{sorces_action}
\end{eqnarray}
is left invariant by both $s$ and $\delta $ operators
\begin{equation}
s\widetilde{S}_{\sigma }=\delta \widetilde{S}_{\sigma }=0\;.  \label{ni}
\end{equation}
Furthermore, it turns out that the original term $S_{\sigma }$ is recovered
from $\widetilde{S}_{\sigma }$ when the external sources attain their
physical values, defined as
\begin{eqnarray}
\left. \bar{X}_{\mu \nu }\right| _{\mathrm{phys}} &\!\!\!\!=\!\!\!\!&\left.
U_{\mu \nu }\right| _{\mathrm{phys}}=-\frac{\sigma ^{2}}{2}\delta _{\mu \nu
}\;,  \nonumber \\
\left. X_{\mu \nu }\right| _{\mathrm{phys}} &\!\!\!\!=\!\!\!\!&\left. \bar{U}%
_{\mu \nu }\right| _{\mathrm{phys}}=\left. Y_{\mu \nu }\right| _{\mathrm{phys%
}}=\left. \bar{Y}_{\mu \nu }\right| _{\mathrm{phys}}=\left. V_{\mu \nu
}\right| _{\mathrm{phys}}=\left. \bar{V}_{\mu \nu }\right| _{\mathrm{phys}%
}=0\;.  \label{pv}
\end{eqnarray}
Thus, we have
\begin{equation}
\left. \widetilde{S}_{\sigma }\right| _{\mathrm{phys}}\to S_{\sigma }\;.
\label{pl}
\end{equation}
The previous equation allows us to introduce a more general action
\begin{equation}
\widetilde{S}_{\mathrm{Local}}=S_{0}+\widetilde{S}_{\sigma }\;,
\label{new_local}
\end{equation}
where $S_{0}$ and $\widetilde{S}_{\sigma }$ are respectively given by $%
\left( \ref{massless}\right) $ and $\left( \ref{sorces_action}\right) $,
which is left invariant by both $s$ and $\delta $ operators,

\begin{equation}
s\widetilde{S}_{\mathrm{Local}}=\delta \widetilde{S}_{\mathrm{Local}}=0\;,
\label{fg}
\end{equation}
while reducing to the action $S_{\mathrm{Local}}$, eq.$\left( \ref{local}%
\right) $, when the sources attain their physical values

\begin{equation}
\left. \widetilde{S}_{\mathrm{Local}}\right| _{\mathrm{phys}}\to S_{_{%
\mathrm{Local}}}\;.  \label{emb}
\end{equation}
We see thus that the action $S_{_{\mathrm{Local}}}$ has been embedded in a
more general action, $\widetilde{S}_{\mathrm{Local}}$, exhibiting exact
invariance. Moreover, $\widetilde{S}_{\mathrm{Local}}$ turns out to display
a further global symmetry $U(4)$:
\begin{equation}
\mathcal{Q}_{\mu \nu }\,\widetilde{S}_{\mathrm{Local}}=0\;,  \label{uu4}
\end{equation}
where
\begin{eqnarray}
\mathcal{Q}_{\mu \nu } &\equiv &\int d^{4}\!x\,\left( B_{\mu }^{a}\frac{%
\delta }{\delta B_{\nu }^{a}}-\bar{B}_{\nu }^{a}\frac{\delta }{\delta \bar{B}%
_{\mu }^{a}}+G_{\mu }^{a}\frac{\delta }{\delta G_{\nu }^{a}}-\bar{G}_{\nu
}^{a}\frac{\delta }{\delta \bar{G}_{\mu }^{a}}+X_{\mu \sigma }\frac{\delta }{%
\delta X_{\nu \sigma }}-\bar{X}_{\nu \sigma }\frac{\delta }{\delta \bar{X}%
_{\mu \sigma }}\right.   \nonumber \\
&&\left. +Y_{\mu \sigma }\frac{\delta }{\delta Y_{\nu \sigma }}-\bar{Y}_{\nu
\sigma }\frac{\delta }{\delta \bar{Y}_{\mu \sigma }}+U_{\mu \sigma }\frac{%
\delta }{\delta U_{\nu \sigma }}-\bar{U}_{\nu \sigma }\frac{\delta }{\delta
\bar{U}_{\mu \sigma }}+V_{\mu \sigma }\frac{\delta }{\delta V_{\nu \sigma }}-%
\bar{V}_{\nu \sigma }\frac{\delta }{\delta \bar{V}_{\mu \sigma }}\right) \;.
\label{u4op}
\end{eqnarray}
This symmetry can be associated to a new quantum number whose generator is
the trace of $\mathcal{Q}_{\mu \nu }$, \textit{i.e.}, $\mathcal{Q}_{4}\equiv
\mathcal{Q}_{\mu \mu }$. As already noticed in \cite
{Zwanziger:1989mf,Zwanziger:1992qr,Capri:2005dy,Capri:2006ne}, the existence
of this global invariance allows us to differentiate between the indices
which refer to $U(4)$ and the remaining Lorentz indices. Denoting by $%
i,j,k,...$, the indices corresponding to the $U(4)$ invariance, expressions $%
\left( \ref{uu4}\right) $ and $\left( \ref{u4op}\right) $ can be rewritten
as
\begin{equation}
\mathcal{Q}_{ij}\,\widetilde{S}_{\mathrm{Local}}=0\;,  \label{u4r}
\end{equation}
\begin{eqnarray}
\mathcal{Q}_{ij} &\equiv &\int d^{4}\!x\,\left( B_{i}^{a}\frac{\delta }{%
\delta B_{j}^{a}}-\bar{B}_{j}^{a}\frac{\delta }{\delta \bar{B}_{i}^{a}}%
+G_{i}^{a}\frac{\delta }{\delta G_{j}^{a}}-\bar{G}_{j}^{a}\frac{\delta }{%
\delta \bar{G}_{i}^{a}}+X_{i\mu }\frac{\delta }{\delta X_{j\mu }}-\bar{X}%
_{j\mu }\frac{\delta }{\delta \bar{X}_{i\mu }}\right.   \nonumber \\
&&\left. +Y_{i\mu }\frac{\delta }{\delta Y_{j\mu }}-\bar{Y}_{j\mu }\frac{%
\delta }{\delta \bar{Y}_{i\mu }}+U_{i\mu }\frac{\delta }{\delta U_{j\mu }}-%
\bar{U}_{j\mu }\frac{\delta }{\delta \bar{U}_{i\mu }}+V_{i\mu }\frac{\delta
}{\delta V_{j\mu }}-\bar{V}_{j\mu }\frac{\delta }{\delta \bar{V}_{i\mu }}%
\right) \;,  \label{u4opr}
\end{eqnarray}
and
\begin{equation}
\widetilde{S}_{\mathrm{Local}}=S_{\mathrm{YM}}+S_{\mathrm{gf}}+S_{\mathrm{aux%
}}+\widetilde{S}_{\sigma }\;,  \label{ar}
\end{equation}
with $S_{\mathrm{aux}}$, $\widetilde{S}_{\sigma }$ given by
\begin{eqnarray}
S_{\mathrm{aux}} &=&\frac{1}{2}\int d^{4}\!x\,\left( \bar{B}_{i}^{a}D_{\mu
}^{ab}D_{\mu }^{bc}B_{i}^{c}-\bar{G}_{i}^{a}D_{\mu }^{ab}D_{\mu
}^{bc}G_{i}^{c}\;\right) \;,  \nonumber \\
\widetilde{S}_{\sigma } &=&\int d^{4}\!x\,\left[ -\bar{X}_{i\mu
}B_{i}^{a}A_{\mu }^{a}+Y_{i\mu }\bar{G}_{i}^{a}A_{\mu }^{a}+U_{i\mu }\bar{B}%
_{i}^{a}A_{\mu }^{a}+\bar{V}_{i\mu }G_{i}^{a}A_{\mu }^{a}\right.   \nonumber
\\
&&\left. -\left( \bar{Y}_{i\mu }B_{i}^{a}-X_{i\mu }\bar{G}_{i}^{a}-V_{i\mu }%
\bar{B}_{i}^{a}-\bar{U}_{i\mu }G_{i}^{a}\right) \partial _{\mu
}c^{a}\;\right] \;.  \label{i-notation}
\end{eqnarray}

\subsection{Identification of the final complete classical action}

We can now identify the complete classical action to start with. To this
purpose, the action $\widetilde{S}_{\mathrm{Local}}$ has to be supplemented
by three additional terms given, respectively, by
\begin{eqnarray}
S_{\mathrm{ext}} &=&s\int d^{4}\!x\,\left( -\Omega _{\mu }^{a}A_{\mu
}^{a}+L^{a}c^{a}\right) +s\delta \int d^{4}\!x\,\left( \bar{N}%
_{i}^{a}B_{i}^{a}+M_{i}^{a}\bar{G}_{i}^{a}\right)   \nonumber \\
&=&\int d^{4}\!x\,\left( -\Omega _{\mu }^{a}D_{\mu }^{ab}c^{b}+\frac{1}{2}%
f^{abc}L^{a}c^{b}c^{c}+gf^{abc}\bar{M}%
_{i}^{a}c^{b}B_{i}^{c}+gf^{abc}M_{i}^{a}c^{b}\bar{B}_{i}^{c}\right.
\nonumber \\
&&\left. +gf^{abc}\bar{N}_{i}^{a}c^{b}G_{i}^{c}+gf^{abc}N_{i}^{a}c^{b}\bar{G}%
_{i}^{c}\right) \;,  \label{s1}
\end{eqnarray}
\begin{equation}
S_{\lambda }=\delta \int d^{4}\!x\,\frac{\lambda ^{abcd}}{16}\,\bar{G}%
_{i}^{a}B_{i}^{b}\left( \bar{B}_{j}^{c}B_{j}^{d}-\bar{G}_{j}^{c}G_{j}^{d}%
\right) =\int d^{4}\!x\,\frac{\lambda ^{abcd}}{16}\left( \bar{B}%
_{i}^{a}B_{i}^{b}-\bar{G}_{i}^{a}G_{i}^{b}\right) \left( \bar{B}%
_{j}^{c}B_{j}^{d}-\bar{G}_{j}^{c}G_{j}^{d}\right) \;,  \label{lambda}
\end{equation}
and
\begin{equation}
S_{\mathrm{\zeta }}=s\delta \left( \zeta \int d^{4}\!x\,\,\bar{V}_{i\mu
}V_{i\mu }\right) =\zeta \int d^{4}\!x\,\left( \bar{X}_{i\mu }U_{i\mu }-\bar{%
V}_{i\mu }Y_{i\mu }\right) \;.  \label{qs}
\end{equation}
Let us analyze each term separately. The first one, eq.$\left( \ref{s1}%
\right) $, is needed in order to take into account the nonlinear $BRST\;$%
transformations of the fields, see eq.$\left( \ref{brst}\right) $. In this
term we have introduced new external sources $\Omega _{\mu }^{a},$ $L^{a},$ $%
M_{i}^{a},$ $\bar{M}_{i}^{a},$ $N_{i}^{a},$ $\bar{N}_{i}^{a}\;,$ which
transform as
\begin{equation}
s\Omega _{\mu }^{a}=sL^{a}=sM_{i}^{a}=s\bar{M}_{i}^{a}=sN_{i}^{a}=s\bar{N}%
_{i}^{a}=0\;,  \label{next}
\end{equation}
and
\begin{eqnarray}
\delta \bar{N}_{i}^{a} &\!\!\!=\!\!\!&-\bar{M}_{i}^{a}\;,\qquad \delta \bar{M%
}_{i}^{a}=0\;,  \nonumber \\
\delta M_{i}^{a} &\!\!\!=\!\!\!&N_{i}^{a}\;,\qquad \delta N_{i}^{a}=0\;,
\nonumber \\
\delta \Omega _{\mu }^{a} &\!\!\!=\!\!\!&0\;,\qquad \delta L^{a}=0\;,
\label{dext}
\end{eqnarray}
ensuring both $s$ and $\delta $ invariance of $S_{\mathrm{ext}}$. The second
term, $S_{\lambda }$, is a quartic term in the auxiliary fields, allowed by
power counting. As such it has to be introduced from the beginning. The $%
\delta $ invariance of $S_{\lambda }$ is manifest. Its $BRST$ invariance is
achieved by demanding that the quartic coupling $\lambda ^{abcd}$ is an
invariant tensor in the adjoint representation, namely
\begin{equation}
f^{man}\lambda ^{mbcd}+f^{mbn}\lambda ^{amcd}+f^{mcn}\lambda
^{abmd}+f^{mdn}\lambda ^{abcm}=0\;.  \label{lambda1}
\end{equation}
Also, from expression $\left( \ref{lambda}\right) $ one easily infers that $%
\lambda ^{abcd}$ possesses the following symmetry properties

\begin{equation}
\lambda ^{abcd}=\lambda ^{cdab}=\lambda ^{bacd}\;.  \label{lambda2}
\end{equation}
Finally, the third term, $S_{\mathrm{\zeta }}$, obviously invariant under $s$
and $\delta $, contains terms which depend only on the external sources.
This term is allowed by power counting, being in fact needed for the
renormalizability of the model. The parameter $\zeta $ in expression $\left(
\ref{qs}\right) $ is a constant dimensionless parameter. \newline
\newline
Therefore, for the complete starting classical action $\Sigma $, we obtain
\begin{eqnarray}
\Sigma  &=&S_{\mathrm{YM}}+S_{\mathrm{gf}}+\widetilde{S}_{\sigma }+S_{%
\mathrm{ext}}+S_{\lambda }+S_{\mathrm{\zeta }}  \nonumber \\
&=&S_{\mathrm{YM}}+\int d^{4}\!x\,\left[ b^{a}\partial _{\mu }A_{\mu }^{a}+%
\bar{c}^{a}\partial _{\mu }D_{\mu }^{ab}c^{b}+\frac{1}{2}\bar{B}%
_{i}^{a}D_{\mu }^{ab}D_{\mu }^{bc}B_{i}^{c}-\frac{1}{2}\bar{G}_{i}^{a}D_{\mu
}^{ab}D_{\mu }^{bc}G_{i}^{c}-\bar{X}_{i\mu }B_{i}^{a}A_{\mu }^{a}\right.
\nonumber \\
&&+Y_{i\mu }\bar{G}_{i}^{a}A_{\mu }^{a}+U_{i\mu }\bar{B}_{i}^{a}A_{\mu }^{a}+%
\bar{V}_{i\mu }G_{i}^{a}A_{\mu }^{a}-\left( \bar{Y}_{i\mu }B_{i}^{a}-X_{i\mu
}\bar{G}_{i}^{a}-V_{i\mu }\bar{B}_{i}^{a}-\bar{U}_{i\mu }G_{i}^{a}\right)
\partial _{\mu }c^{a}  \nonumber \\
&&+\zeta \left( \bar{X}_{i\mu }U_{i\mu }-\bar{V}_{i\mu }Y_{i\mu }\right)
-\Omega _{\mu }^{a}D_{\mu }^{ab}c^{b}+\frac{g}{2}%
f^{abc}L^{a}c^{b}c^{c}+gf^{abc}\bar{M}%
_{i}^{a}c^{b}B_{i}^{c}+gf^{abc}M_{i}^{a}c^{b}\bar{B}_{i}^{c}  \nonumber \\
&&\left. +gf^{abc}\bar{N}_{i}^{a}c^{b}G_{i}^{c}+gf^{abc}N_{i}^{a}c^{b}\bar{G}%
_{i}^{c}+\frac{\lambda ^{abcd}}{16}\,\left( \bar{B}_{i}^{a}B_{i}^{b}-\bar{G}%
_{i}^{a}G_{i}^{b}\right) \left( \bar{B}_{j}^{c}B_{j}^{d}-\bar{G}%
_{j}^{c}G_{j}^{d}\right) \right] \;.  \label{start}
\end{eqnarray}
Let us also display, for further use, all quantum numbers of the fields and
sources, given in tables \ref{table1} and \ref{table2}.

\begin{table}[t]
\centering
\begin{tabular}{lcccccccc}
\hline\hline
& $A$ & $b$ & $\bar c$ & $c$ & $B$ & $\bar{B}$ & $G$ & $\bar{G}%
\phantom{\Bigl|}$ \\ \hline
dimension & $1$ & $2$ & $2$ & $0$ & $1$ & $1$ & $1$ & $1$ \\
ghost number & $0$ & $0$ & $-1$ & $1$ & $0$ & $0$ & $0$ & $0$ \\
$\mathcal{N}_{\mathrm{f}}$ number & $0$ & $0$ & $0$ & $0$ & $0$ & $0$ & $1$
& $-1$ \\
$\mathcal{Q}_4$-charge & $0$ & $0$ & $0$ & $0$ & $1$ & $-1$ & $1$ & $-1$ \\
\hline\hline
\end{tabular}
\caption{The quantum numbers of the fields.}
\label{table1}
\end{table}

\begin{table}[t]
\centering
\begin{tabular}{lcccccccccccccc}
\hline\hline
& $\Omega$ & $L$ & $X$ & $\bar{X}$ & $Y$ & $\bar{Y}$ & $U$ & $\bar{U}$ & $V$
& $\bar{V}$ & $M$ & $\bar{M}$ & $N$ & $\bar{N}\phantom{\Bigl|}$ \\ \hline
dimension & $3$ & $4$ & $2$ & $2$ & $2$ & $2$ & $2$ & $2$ & $2$ & $2$ & $3$
& $3$ & $3$ & $3$ \\
ghost number & $-1$ & $-2$ & $-1$ & $0$ & $0$ & $-1$ & $0$ & $-1$ & $-1$ & $%
0 $ & $-1$ & $-1$ & $-1$ & $-1$ \\
$\mathcal{N}_{\mathrm{f}}$ number & $0$ & $0$ & $1$ & $0$ & $1$ & $0$ & $0$
& $-1$ & $0$ & $-1$ & $0$ & $0$ & $1$ & $-1$ \\
$\mathcal{Q}_4$-charge & $0$ & $0$ & $1$ & $-1$ & $1$ & $-1$ & $1$ & $-1$ & $%
1$ & $-1$ & $1$ & $-1$ & $1$ & $-1$ \\ \hline\hline
\end{tabular}
\caption{The quantum numbers of the sources.}
\label{table2}
\end{table}

\section{Ward identities.}

The action $\left( \text{\ref{start}}\right) $ enjoys a large set of Ward
identities. In fact, it turns out that $\Sigma $ fulfills:

\begin{itemize}
\item  {The Slavnov-Taylor identity }
\begin{equation}
\mathcal{S}(\Sigma )=0\;,  \label{st}
\end{equation}
{\
\begin{eqnarray}
\mathcal{S}(\Sigma ) &\equiv &\int d^{4}\!x\,\left( \frac{\delta \Sigma }{%
\delta \Omega _{\mu }^{a}}\frac{\delta \Sigma }{\delta A_{\mu }^{a}}+\frac{%
\delta \Sigma }{\delta L^{a}}\frac{\delta \Sigma }{\delta c^{a}}+\frac{%
\delta \Sigma }{\delta \bar{M}_{i}^{a}}\frac{\delta \Sigma }{\delta B_{i}^{a}%
}+\frac{\delta \Sigma }{\delta M_{i}^{a}}\frac{\delta \Sigma }{\delta \bar{B}%
_{i}^{a}}+\frac{\delta \Sigma }{\delta \bar{N}_{i}^{a}}\frac{\delta \Sigma }{%
\delta G_{i}^{a}}\right.   \nonumber \\
&&\left. +\frac{\delta \Sigma }{\delta N_{i}^{a}}\frac{\delta \Sigma }{%
\delta \bar{G}_{i}^{a}}+b^{a}\frac{\delta \Sigma }{\delta \bar{c}^{a}}%
+Y_{i\mu }\frac{\delta \Sigma }{\delta X_{i\mu }}+\bar{X}_{i\mu }\frac{%
\delta \Sigma }{\delta \bar{Y}_{i\mu }}+U_{i\mu }\frac{\delta \Sigma }{%
\delta V_{i\mu }}+\bar{V}_{i\mu }\frac{\delta \Sigma }{\delta \bar{U}_{i\mu }%
}\right) \;,  \label{slavnov}
\end{eqnarray}
}

\item  {The Landau gauge-fixing condition
\begin{equation}
\frac{\delta \Sigma }{\delta b^{a}}=\partial _{\mu }A_{\mu }^{a}\;,
\label{gauge-fixing}
\end{equation}
}

\item  {The anti-ghost equation
\begin{equation}
\frac{\delta \Sigma }{\delta \bar{c}^{a}}+\partial _{\mu }\frac{\delta
\Sigma }{\delta \Omega _{\mu }^{a}}=0\;,  \label{anti-ghost}
\end{equation}
}

\item  {The ghost equation
\begin{equation}
\mathcal{G}^{a}(\Sigma )=\Delta _{\mathrm{class}}^{a}\;,  \label{ghost}
\end{equation}
where
\begin{equation}
\mathcal{G}^{a}\equiv \int d^{4}\!x\,\left( \frac{\delta \Sigma }{\delta
c^{a}}+gf^{abc}\bar{c}^{b}\frac{\delta \Sigma }{\delta b^{c}}\right) \;,
\label{ghop}
\end{equation}
and
\begin{equation}
\Delta _{\mathrm{class}}^{a}=\int d^{4}\!x\,gf^{abc}\left( \Omega _{\mu
}^{b}A_{\mu }^{c}-L^{b}c^{c}+\bar{M}_{i}^{b}B_{i}^{c}+M_{i}^{b}\bar{B}%
_{i}^{c}-\bar{N}_{i}^{b}G_{i}^{c}-N_{i}^{b}\bar{G}_{i}^{c}\right) \;.
\label{ghb}
\end{equation}
}

Notice that the breaking term $\Delta _{\mathrm{class}}^{a}$ is linear in
the quantum fields. It is thus a classical breaking, not affected by the
quantum corrections \cite{Piguet:1995er}.

\item  {The ghost number Ward identity
\begin{equation}
\mathcal{N}_{\mathrm{gh}}(\Sigma )=0\;,  \label{ghost_number}
\end{equation}
with
\begin{eqnarray}
\mathcal{N}_{\mathrm{gh}} &\equiv &\int d^{4}\!x\,\left( c^{a}\frac{\delta }{%
\delta c^{a}}-\bar{c}^{a}\frac{\delta }{\delta \bar{c}^{a}}-\bar{Y}_{i\mu }%
\frac{\delta }{\delta \bar{Y}_{i\mu }}-X_{i\mu }\frac{\delta }{\delta
X_{i\mu }}-V_{i\mu }\frac{\delta }{\delta V_{i\mu }}-\bar{U}_{i\mu }\frac{%
\delta }{\delta \bar{U}_{i\mu }}\right.   \nonumber \\
&&\left. -\Omega _{\mu }^{a}\frac{\delta }{\delta \Omega _{\mu }^{a}}-2L^{a}%
\frac{\delta }{\delta L^{a}}-\bar{M}_{i}^{a}\frac{\delta }{\delta \bar{M}%
_{i}^{a}}-M_{i}^{a}\frac{\delta }{\delta M_{i}^{a}}-\bar{N}_{i}^{a}\frac{%
\delta }{\delta \bar{N}_{i}^{a}}-N_{i}^{a}\frac{\delta }{\delta N_{i}^{a}}%
\right) \;,
\end{eqnarray}
}

\item  {The Ward identity corresponding to the fermionic invariance }$\left(
\text{\ref{nf}}\right) ${\
\begin{equation}
\mathcal{N}_{\mathrm{f}}(\Sigma )=0\;,  \label{nfc}
\end{equation}
where }$\mathcal{N}_{\mathrm{f}}$ stands for the operator{\
\begin{eqnarray}
\mathcal{N}_{\mathrm{f}} &\equiv &\int d^{4}\!x\,\left( G_{i}^{a}\frac{%
\delta }{\delta G_{i}^{a}}-\bar{G}_{i}^{a}\frac{\delta }{\delta \bar{G}%
_{i}^{a}}+Y_{i\mu }\frac{\delta }{\delta Y_{i\mu }}+X_{i\mu }\frac{\delta }{%
\delta X_{i\mu }}\right.   \nonumber \\
&&\left. -\bar{V}_{i\mu }\frac{\delta }{\delta \bar{V}_{i\mu }}-\bar{U}%
_{i\mu }\frac{\delta }{\delta \bar{U}_{i\mu }}+\bar{N}_{i}^{a}\frac{\delta }{%
\delta \bar{N}_{i}^{a}}-N_{i}^{a}\frac{\delta }{\delta N_{i}^{a}}\right) \;.
\label{nfop}
\end{eqnarray}
}

\item  {The global $U(4)$ invariance
\begin{equation}
\mathcal{Q}_{ij}(\Sigma )=0\;,  \label{u4}
\end{equation}
where
\begin{eqnarray}
\mathcal{Q}_{ij} &\equiv &\int d^{4}\!x\,\left( B_{i}^{a}\frac{\delta }{%
\delta B_{j}^{a}}-\bar{B}_{j}^{a}\frac{\delta }{\delta \bar{B}_{i}^{a}}%
+G_{i}^{a}\frac{\delta }{\delta G_{j}^{a}}-\bar{G}_{j}^{a}\frac{\delta }{%
\delta \bar{G}_{i}^{a}}+X_{i\mu }\frac{\delta }{\delta X_{j\mu }}-\bar{X}%
_{j\mu }\frac{\delta }{\delta \bar{X}_{i\mu }}\right.   \nonumber \\
&&+Y_{i\mu }\frac{\delta }{\delta Y_{j\mu }}-\bar{Y}_{j\mu }\frac{\delta }{%
\delta \bar{Y}_{i\mu }}+U_{i\mu }\frac{\delta }{\delta U_{j\mu }}-\bar{U}%
_{j\mu }\frac{\delta }{\delta \bar{U}_{i\mu }}+V_{i\mu }\frac{\delta }{%
\delta V_{j\mu }}-\bar{V}_{j\mu }\frac{\delta }{\delta \bar{V}_{i\mu }}
\nonumber \\
&&\left. +M_{i}^{a}\frac{\delta }{\delta M_{j}^{a}}-\bar{M}_{j}^{a}\frac{%
\delta }{\delta \bar{M}_{i}^{a}}+N_{i}^{a}\frac{\delta }{\delta N_{j}^{a}}-%
\bar{N}_{j}^{a}\frac{\delta }{\delta \bar{N}_{i}^{a}}\right) \;,  \label{qij}
\end{eqnarray}
}

\item  {The rigid symmetries
\begin{equation}
\mathcal{R}_{ij}^{(A)}(\Sigma )=0\;,\qquad (A=1,2,3,4)\;,  \label{Ra}
\end{equation}
where
\begin{eqnarray}
\mathcal{R}_{ij}^{(1)} &\!\!\!\!\equiv \!\!\!\!&\int d^{4}\!x\,\left(
B_{i}^{a}\frac{\delta }{\delta G_{j}^{a}}-\bar{G}_{j}^{a}\frac{\delta }{%
\delta \bar{B}_{i}^{a}}-V_{i\mu }\frac{\delta }{\delta X_{j\mu }}-\bar{V}%
_{j\mu }\frac{\delta }{\delta \bar{X}_{i\mu }}+U_{i\mu }\frac{\delta }{%
\delta Y_{j\mu }}+\bar{U}_{j\mu }\frac{\delta }{\delta \bar{Y}_{i\mu }}%
+M_{i}^{a}\frac{\delta }{\delta N_{j}^{a}}-\bar{N}_{j}^{a}\frac{\delta }{%
\delta \bar{M}_{i}^{a}}\right) \;,  \nonumber \\
\mathcal{R}_{ij}^{(2)} &\!\!\!\!\equiv \!\!\!\!&\int d^{4}\!x\,\left( \bar{B}%
_{i}^{a}\frac{\delta }{\delta \bar{G}_{j}^{a}}+G_{j}^{a}\frac{\delta }{%
\delta B_{i}^{a}}+\bar{X}_{i\mu }\frac{\delta }{\delta \bar{V}_{j\mu }}-{X}%
_{j\mu }\frac{\delta }{\delta {V}_{i\mu }}-\bar{Y}_{i\mu }\frac{\delta }{%
\delta \bar{U}_{j\mu }}+{Y}_{j\mu }\frac{\delta }{\delta {U}_{i\mu }}-\bar{M}%
_{i}^{a}\frac{\delta }{\delta \bar{N}_{j}^{a}}+{N}_{j}^{a}\frac{\delta }{%
\delta {M}_{i}^{a}}\right) \;,  \nonumber \\
\mathcal{R}_{ij}^{(3)} &\!\!\!\!\equiv \!\!\!\!&\int d^{4}\!x\,\left( \bar{B}%
_{i}^{a}\frac{\delta }{\delta G_{j}^{a}}-\bar{G}_{j}^{a}\frac{\delta }{%
\delta {B}_{i}^{a}}+\bar{Y}_{i\mu }\frac{\delta }{\delta X_{j\mu }}-\bar{X}%
_{i\mu }\frac{\delta }{\delta {Y}_{j\mu }}+\bar{V}_{j\mu }\frac{\delta }{%
\delta U_{i\mu }}-\bar{U}_{j\mu }\frac{\delta }{\delta {V}_{i\mu }}+\bar{M}%
_{i}^{a}\frac{\delta }{\delta N_{j}^{a}}+\bar{N}_{j}^{a}\frac{\delta }{%
\delta {M}_{i}^{a}}\right) \;,  \nonumber \\
\mathcal{R}_{ij}^{(4)} &\!\!\!\!\equiv \!\!\!\!&\int d^{4}\!x\,\left(
B_{i}^{a}\frac{\delta }{\delta \bar{G}_{j}^{a}}+{G}_{j}^{a}\frac{\delta }{%
\delta \bar{B}_{i}^{a}}-U_{i\mu }\frac{\delta }{\delta \bar{V}_{j\mu }}-{Y}%
_{j\mu }\frac{\delta }{\delta \bar{X}_{i\mu }}+V_{i\mu }\frac{\delta }{%
\delta \bar{U}_{j\mu }}+{X}_{j\mu }\frac{\delta }{\delta \bar{Y}_{i\mu }}%
-M_{i}^{a}\frac{\delta }{\delta \bar{N}_{j}^{a}}+{N}_{j}^{a}\frac{\delta }{%
\delta \bar{M}_{i}^{a}}\right) \;.  \nonumber \\
&&  \label{rs}
\end{eqnarray}
}
\end{itemize}

\noindent As we shall see in the next section, this set of Ward identities
will enable us to prove the renormalizability of the complete action $\left(
\text{\ref{start}}\right) $.

\section{Algebraic characterization of the invariant counterterm and
renormalizability}

Having established all Ward identities fulfilled by the complete action, eq.%
\eqref{start}, we can now turn our attention to the characterization of the
most general invariant counterterm $\Sigma _{\mathrm{CT}}$. Following the
algebraic renormalization procedure \cite{Piguet:1995er}, $\Sigma _{\mathrm{%
CT}}$ has to be an integrated local polynomial in the fields and sources
with dimension bounded by four, with vanishing ghost and $\mathcal{N}_{%
\mathrm{f}}$ numbers as well as $\mathcal{Q}_{4}$-charge, and obeying the
following constraints
\begin{eqnarray}
\mathcal{B}_{\Sigma }\,\Sigma _{\mathrm{CT}} &=&0\;,  \nonumber \\
\frac{\delta }{\delta b^{a}}\,\Sigma _{\mathrm{CT}} &=&0\;,  \nonumber \\
\left( \frac{\delta }{\delta \bar{c}^{a}}+\partial _{\mu }\frac{\delta }{%
\delta \Omega _{\mu }^{a}}\right) \,\Sigma _{\mathrm{CT}} &=&0\;,  \nonumber
\\
\mathcal{G}^{a}\,\Sigma _{\mathrm{CT}} &=&0\;,  \nonumber \\
\mathcal{N}_{\mathrm{gh}}\,\Sigma _{\mathrm{CT}} &=&0\;,  \nonumber \\
\mathcal{N}_{\mathrm{f}}\,\Sigma _{\mathrm{CT}} &=&0\;,  \nonumber \\
\mathcal{Q}_{ij}\,\Sigma _{\mathrm{CT}} &=&0\;,  \nonumber \\
\mathcal{R}_{ij}^{(A)}\,\Sigma _{\mathrm{CT}} &=&0\;,  \label{cst}
\end{eqnarray}
where $\mathcal{B}_{\Sigma }$ is the nilpotent linearized Slavnov-Taylor
operator
\begin{eqnarray}
\mathcal{B}_{\Sigma } &\equiv &\int d^{4}\!x\,\left( \frac{\delta \Sigma }{%
\delta \Omega _{\mu }^{a}}\frac{\delta }{\delta A_{\mu }^{a}}+\frac{\delta
\Sigma }{\delta A_{\mu }^{a}}\frac{\delta }{\delta \Omega _{\mu }^{a}}+\frac{%
\delta \Sigma }{\delta L^{a}}\frac{\delta }{\delta c^{a}}+\frac{\delta
\Sigma }{\delta c^{a}}\frac{\delta }{\delta L^{a}}+\frac{\delta \Sigma }{%
\delta \bar{M}_{i}^{a}}\frac{\delta }{\delta B_{i}^{a}}+\frac{\delta \Sigma
}{\delta B_{i}^{a}}\frac{\delta }{\delta \bar{M}_{i}^{a}}\right.   \nonumber
\\
&&+\frac{\delta \Sigma }{\delta M_{i}^{a}}\frac{\delta }{\delta \bar{B}%
_{i}^{a}}+\frac{\delta \Sigma }{\delta \bar{B}_{i}^{a}}\frac{\delta }{\delta
M_{i}^{a}}+\frac{\delta \Sigma }{\delta \bar{N}_{i}^{a}}\frac{\delta }{%
\delta G_{i}^{a}}+\frac{\delta \Sigma }{\delta G_{i}^{a}}\frac{\delta }{%
\delta \bar{N}_{i}^{a}}+\frac{\delta \Sigma }{\delta N_{i}^{a}}\frac{\delta
}{\delta \bar{G}_{i}^{a}}+\frac{\delta \Sigma }{\delta \bar{G}_{i}^{a}}\frac{%
\delta }{\delta N_{i}^{a}}  \nonumber \\
&&\left. +b^{a}\frac{\delta }{\delta \bar{c}^{a}}+Y_{i\mu }\frac{\delta }{%
\delta X_{i\mu }}+\bar{X}_{i\mu }\frac{\delta }{\delta \bar{Y}_{i\mu }}%
+U_{i\mu }\frac{\delta }{\delta V_{i\mu }}+\bar{V}_{i\mu }\frac{\delta }{%
\delta \bar{U}_{i\mu }}\right) \;,  \label{nbrst}
\end{eqnarray}
\begin{equation}
\mathcal{B}_{\Sigma }\mathcal{B}_{\Sigma }=0\;.  \label{np}
\end{equation}
After a rather lengthy analysis, the most general allowed counterterm $%
\Sigma _{\mathrm{CT}}$ compatible with all Ward identities is found to be
\begin{eqnarray}
\Sigma _{\mathrm{CT}} &=&a_{0}\,S_{\mathrm{YM}}+\int d^{4}\!x\,\left\{
a_{1}\,A_{\mu }^{a}\frac{\delta S_{\mathrm{YM}}}{\delta A_{\mu }^{a}}%
+a_{1}\,\left( \Omega _{\mu }^{a}+\partial _{\mu }\bar{c}^{a}\right)
\partial _{\mu }c^{a}+\frac{a_{2}}{2}\,\left( \bar{B}_{i}^{a}\partial
^{2}B_{i}^{a}-\bar{G}_{i}^{a}\partial ^{2}G_{i}^{a}\right) \right.
\nonumber \\
&&-(a_{1}+a_{2})\left[ \frac{g}{2}f^{abc}\left( \bar{B}_{i}^{a}B_{i}^{b}-%
\bar{G}_{i}^{a}G_{i}^{b}\right) \partial _{\mu }A_{\mu }^{c}+gf^{abc}\left(
\bar{B}_{i}^{a}\partial _{\mu }B_{i}^{b}-\bar{G}_{i}^{a}\partial _{\mu
}G_{i}^{b}\right) A_{\mu }^{c}\right]   \nonumber \\
&&+(2a_{1}+a_{2})\,\frac{g^{2}}{2}f^{acd}f^{cbe}\left( \bar{B}%
_{i}^{a}B_{i}^{b}-\bar{G}_{i}^{a}G_{i}^{b}\right) A_{\mu }^{d}A_{\mu
}^{e}-(a_{1}+a_{3})\left( \bar{X}_{i\mu }B_{i}^{a}-\bar{V}_{i\mu
}G_{i}^{a}\right.   \nonumber \\
&&\left. -Y_{i\mu }\bar{G}_{i}^{a}-U_{i\mu }\bar{B}_{i}^{a}\right) A_{\mu
}^{a}+a_{3}\left( X_{i\mu }\bar{G}_{i}^{a}+V_{i\mu }\bar{B}_{i}^{a}-\bar{Y}%
_{i\mu }B_{i}^{a}+\bar{U}_{i\mu }G_{i}^{a}\right) \partial _{\mu }c^{a}
\nonumber \\
&&+\left[ (2a_{2}+a_{4})\frac{\lambda ^{abcd}}{16}+a_{4}\frac{\mathcal{N}%
^{abcd}}{16}\right] \left( \bar{B}_{i}^{a}B_{i}^{b}-\bar{G}%
_{i}^{a}G_{i}^{b}\right) \left( \bar{B}_{j}^{c}B_{j}^{d}-\bar{G}%
_{j}^{c}G_{j}^{d}\right)   \nonumber \\
&&\left. +a_{5}\,\zeta \left( \bar{X}_{i\mu }U_{i\mu }-\bar{V}_{i\mu
}Y_{i\mu }\right) \right\} \;.  \label{counterterm}
\end{eqnarray}
In the last expression the coefficients $a_{k}$, $k=0,\dots ,5$, are free
parameters and $\mathcal{N}^{abcd}$ is an invariant tensor with the same
properties of $\lambda ^{abcd},$ eqs.$\left( \ref{lambda1}\right) ,\left(
\ref{lambda2}\right) $. As discussed in \cite{Capri:2005dy,Capri:2006ne},
the tensor $\mathcal{N}^{abcd}$ represents the contribution of quartic
counterterms whose group structure does not allow to express them directly
in terms of $\lambda ^{abcd}$. \newline
It remains now to show that the invariant counterterm $\left( \ref
{counterterm}\right) $ can be reabsorbed through a redefinition of the
parameters, fields and sources of the classical starting action $\Sigma $,
according to
\begin{eqnarray}
\Phi _{0} &=&Z_{\Phi }^{1/2}\,\Phi \;,  \nonumber \\
J_{0} &=&Z_{J}\,J\;,  \nonumber \\
\lambda _{0}^{abcd} &=&Z_{\lambda }\,\lambda ^{abcd}+\mathcal{Z}^{abcd}\;,
\label{rn1}
\end{eqnarray}
where
\begin{eqnarray}
\Phi  &\equiv &(A,b,c,\bar{c},B,\bar{B},G,\bar{G})\;,  \nonumber \\
J &\equiv &(g,\zeta ,\Omega ,L,X,\bar{X},Y,\bar{Y},U,\bar{U},V,\bar{V},M,%
\bar{M},N,\bar{N})\;,  \label{rn2}
\end{eqnarray}
so that
\begin{equation}
\Sigma (\Phi _{0},J_{0},\lambda _{0}^{abcd})=\Sigma (\Phi ,J,\lambda
^{abcd})+\varepsilon \,\Sigma _{\mathrm{CT}}+O(\varepsilon ^{2})\;.
\label{rab}
\end{equation}
By direct inspection, the renormalization constants are found to be
\begin{eqnarray}
&&Z_{b}=Z_{L}^{2}=Z_{A}\;,  \nonumber \\
&&Z_{c}=Z_{\bar{c}}=Z_{\Omega }^{2}=Z_{g}^{-1}Z_{A}^{-1/2}\;,  \nonumber \\
&&Z_{G}=Z_{\bar{G}}=Z_{\bar{B}}=Z_{B}\;,  \nonumber \\
&&Z_{\bar{X}}=Z_{Y}=Z_{\bar{V}}=Z_{U}\;,  \nonumber \\
&&Z_{X}=Z_{V}=Z_{\bar{Y}}=Z_{\bar{U}}=Z_{g}^{-1}Z_{A}^{1/2}Z_{U}\;,
\nonumber \\
&&Z_{M}=Z_{\bar{M}}=Z_{N}=Z_{\bar{N}}=Z_{A}^{1/2}Z_{B}^{-1/2}\;,  \label{zd1}
\end{eqnarray}
with
\begin{eqnarray}
Z_{A} &=&1+\varepsilon \,(a_{0}+2a_{1})\;,  \nonumber \\
Z_{g} &=&1-\varepsilon \,\frac{a_{0}}{2}\;,  \nonumber \\
Z_{B} &=&1+\varepsilon \,a_{2}\;,  \nonumber \\
Z_{U} &=&1-\frac{\varepsilon }{2}\,(a_{0}+a_{2}-2a_{3})\;,  \nonumber \\
Z_{\lambda } &=&1+\varepsilon \,a_{4}\;,  \nonumber \\
\mathcal{Z}^{abcd} &=&\varepsilon a_{4}\,\mathcal{N}^{abcd}\;,  \nonumber \\
Z_{\zeta } &=&1+\varepsilon \,(a_{0}+a_{2}-2a_{3}+a_{5})\;.  \label{zd2}
\end{eqnarray}
Equations $\left( \text{\ref{rab}}\right) ,\left( \text{\ref{zd1}}\right)
,\left( \text{\ref{zd2}}\right) $ show that the counterterm $\Sigma _{%
\mathrm{CT}}$ can be reabsorbed by means of a redefinition of the fields,
sources and parameters of the starting action $\Sigma $, establishing thus
the renormalizability of the theory.

\section{Conclusion}

Nonlocal operators are known to play an important role in Yang-Mills
theories. For example, in the absence of quarks, the vacuum expectation
value of the Wilson loop proves to be an order parameter for the confining
and deconfining phases of Yang-Mills theories. Also, for a large class of
loops, the Wilson operator is renormalizable. \newline
\newline
In this work we call attention to the existence of a slightly
different class of nonlocal operators which can be added to the
Yang-Mills action, while leading to a local and renormalizable
theory. These features are encoded in the possibility of achieving
a rather simple localization procedure for these operators. The
renormalizability is thus guaranteed due to the symmetry content
of the resulting local theory. This framework has been illustrated
through the example of the nonlocal operator of expression $\left(
\ref{i0}\right) $. Although many aspects related to these
operators deserve a better understanding, let us spot here a few
remarks which might be useful for further investigation.

\begin{itemize}
\item  The first observation is related to the nonlocal character
of these operators which, when added to the Yang-Mills action, can
induce deep modifications on the infrared behavior of the theory.
These operators could thus be useful in order to investigate
nonperturbative aspects of confining theories. This is best
illustraded by the example of Zwanziger's horizon function $\left(
\ref{i1}\right) $, which implements the restriction of the domain
of integration in the Feynman path integral up to the first Gribov
horizon. In particular, the resulting tree level gluon propagator
turns out to be suppressed in the infrared, according to \cite
{Gribov:1977wm,Zwanziger:1989mf,Zwanziger:1992qr}
\begin{equation}
\langle A_{\mu }^{a}(k)A_{\nu }^{b}(-k)\rangle =\delta ^{ab}\frac{k^{2}}{%
k^{4}+\gamma ^{4}}\left( \delta _{\mu \nu }-\frac{k_{\mu }k_{\nu }}{k^{2}}%
\right) \;.  \label{gp}
\end{equation}
This propagator exhibits complex poles, a feature which is interpreted as a
signal of gluon confinement. In other words, the gluon is destabilized by
the presence of the Gribov horizon, so that is does not belong anymore to
the physical spectrum.

\item  A second remark follows by noting that, due to dimensional reasons,
these operators require the introduction of dimensionful parameters, \textit{%
i.e. }$\gamma $ for the horizon function $\left( \ref{i1}\right) $, $m$ for
the nonlocal mass operator $\left( \ref{i2}\right) $, and $\sigma $ for the
expression $\left( \ref{i0}\right) $. This naturally rises the question of
whether these parameters could be generated in a dynamical way, reflecting
the possibility that the corresponding operators might condense, acquiring a
nonvanishing vacuum expectation value. This could result in the lowering of
the vacuum energy of the theory, signalling that the aforementioned
condensates are in fact energetically favoured. This would require that
these parameters are determined in a self-consistent way through suitable
gap equations. Once again, this point can be illustrated by the example of
the horizon term $\left( \ref{i1}\right) $. From \cite
{Gribov:1977wm,Zwanziger:1989mf,Zwanziger:1992qr}, one learns that the
Gribov parameter $\gamma $ is not a free parameter, being determined
self-consistently through the gap equation
\begin{equation}
\frac{\delta \Gamma }{\delta \gamma ^{2}}=\;0\;,  \label{ge}
\end{equation}
where $\Gamma $ stands for the quantum 1PI effective action evaluated with
the Yang-Mills action supplemented by the horizon term $\left( \ref{i1}%
\right) $. Equation $\left( \ref{ge}\right) $ enables one to express $\gamma
$ as a function of the gauge coupling $g$ and of the invariant scale $%
\Lambda _{QCD}$.

\item  Notice that the gap equation $\left( \ref{ge}\right) $ can be seen as
a variational minimizing condition, stating that the quantum action $\Gamma $
depends minimally from $\gamma $. The same variational principle could be
employed in order to investigate the dynamical origin of the gluon mass
parameter $m$ as well as to study the operator $\left( \ref{i0}\right) $. It
is worth mentioning that this variational principle has been in fact already
employed in the study of the dimension two condensate $\left\langle A_{\mu
}^{a}A_{\mu }^{a}\right\rangle $ \cite{Sorella:2006ax} in the Landau gauge
which, due to the relationship
\begin{equation}
-\frac{1}{4}\int d^{4}xF_{\mu \nu }^{a}\frac{1}{D^{2}}F_{\mu \nu }^{a}\;=%
\frac{1}{2}\int d^{4}xA_{\mu }^{a}A_{\mu }^{a}\;+\mathit{\;}\text{\textrm{%
higher\ order\ terms\ }}.  \label{g12}
\end{equation}
can be seen as evidence in favour of the possible existence of a nonvashing
condensate $\left\langle F\frac{1}{D^{2}}F\right\rangle $.

\item  One further aspect to be investigated is whether two or more nonlocal
operators could be simultaneously added to the Yang-Mills action in such a
way that the resulting theory preserves renormalizability. This would
require the absence of possible mixing among the various nonlocal operators
which could jeopardize the renormalizability. Let us mention here that, so
far, this issue has been investigated by considering the inclusion in the
Yang-Mills action of both Zwanziger's horizon function $\left( \ref{i1}%
\right) $ and the nonlocal mass term $\left( \ref{i2}\right) $. Thanks to
the rich symmetry content, it turns out that there is no mixing between the
two operators, so that the resulting local theory can be proven to be
renormalizable \cite{dd}. This result could allow us to investigate the
effects of the gauge invariant nonlocal mass operator $\left( \ref{i2}%
\right) $ in the presence of the Gribov horizon.

\item  Finally, it would be interesting to look at a more systematic way in
order to search for other nonlocal operators which might lead to
renormalizable theories. The inclusion of matter fields could also be
exploited. For instance, the investigation of a possible nonlocal spinor
mass term preserving chiral invariance could be of a certain interest.
\end{itemize}

\section*{Acknowledgments}

The Conselho Nacional de Desenvolvimento Cient\'{i}fico e Tecnol\'{o}gico
(CNPq-Brazil), the SR2-UERJ and the Coordena{\c{c}}{\~{a}}o de Aperfei{\c{c}}%
oamento de Pessoal de N{\'{i}}vel Superior (CAPES) are gratefully
acknowledged for financial support.

\end{document}